\documentclass[a4paper,10pt]{article}
\usepackage[utf8]{inputenc}
\usepackage{authblk}
\usepackage{cite} 
\usepackage[mathscr]{eucal}
\usepackage{amsfonts}
\usepackage{amsmath}
\usepackage{slashed}
\usepackage{amssymb}
\usepackage{yfonts}
\usepackage{xcolor}
\usepackage{xcite}
\begin{document}
\title{Hydrodynamic representation and energy balance for Dirac and Weyl fermions in curved space-times}
\author[1]{Tonatiuh Matos\thanks{tonatiuh.matos@cinvestav.mx}}
\author[1]{Omar Gallegos\thanks{omar.gallegos@cinvestav.mx}}
\author[2]{Pierre-Henri Chavanis\thanks{chavanis@irsamc.ups-tlse.fr}}
\affil[1]{\small{\textit{Departamento de Física, Centro de Investigación y de Estudios Avanzados del IPN, A.P. 14-740, 07000 CDMX, México.}}}
 \affil[2]{\small{\textit{Laboratoire de Physique Th\'eorique, Universit\'e de Toulouse, CNRS, UPS, France}}}

\maketitle
\begin{abstract}
Using a generalized Madelung transformation, we derive the hydrodynamic representation of the Dirac equation in arbitrary curved space-times coupled to an electromagnetic field. We obtain Dirac-Euler equations for fermions involving  a continuity equation and a first integral of the Bernoulli equation. Comparing between the Dirac and Klein-Gordon equations we  obtain the balance equation for fermion particles. We also use the correspondence between fermions and bosons to derive the hydrodynamic representation of the Weyl equation which is a chiral form of the Dirac equation. 
\end{abstract}

\section{Introduction}
\label{section:introduction}

The Standard Model of elementary particles establishes that there exist two kinds of particles, fermions and bosons. In previous works \cite{Matos:2016ryp}\cite{Chavanis:2016shp}, the energy balance for bosons was derived  starting from the general relativistic Klein-Gordon (KG) equation. In the present work, we study a system of fermions described by the Dirac equation in arbitrary curved space-times taking into account electromagnetic effects. We also use the Weyl equation which is a  chiral form of the Dirac equation due to the relationship between the Lie algebras of the symmetry groups for both systems of particles. We give the hydrodynamic representation of the Dirac and Weyl equations for fermions using previous results obtained for boson particles. This representation is built analogously as in quantum mechanics (QM) and as in the bosonic case \cite{Matos:2016ryp}, where it was introduced by the Madelung transformation in order to find an alternative interpretation of a boson system. This interpretation has been very useful in astrophysics \cite{Chavanis:2016shp}. In this article, we extend the previous transformation to the fermionic case, in the same way we pretend to give an alternative interpretation of the femionic systems.\newline

Many examples of fermion particles in strong gravitational fields can be found in nature. Indeed, the curvature of space-time plays an important role in a neutron star, in the early Universe, or in a fermion cloud (e.g. a dark matter halo) in the vicinity of a black hole. We need to develop a general framework  to identify what are the different energy contributions in such systems. In this work we use the geometrical decomposition of the metric in 3+1 slices and the tetrad formalism to study the particle spin in an arbitrary space-time. We define the gamma matrices in curved space-times and derive the generalized Dirac and Weyl equations. Then, using the Madelung transformation, we introduce a hydrodynamic representation of the Dirac and Weyl spinors. This hydrodynamic representation can help us to describe the fermionic system in a general framework. We can highlight that this description is convenient because it is easier to make a physical interpretation, since the hydrodynamic representation is given in some variable such as number of particles, speed, potential or energy. In fact, a non-obvious result is the energy balance equation, which is the first law of thermodynamics, which comes from the Dirac equation with the Madelung transformation for spinors. Although the equations obtained from this representation are more complicated than in the usual way, it can help us to have a closer answer for interpretations of quantum theory, for example, the de Broglie-Bohm interpretation\cite{Bohm:1951xw,Bohm:1951xx,Bohm:1966sei}. In addition, we can compare the hydrodynamics and energy balance in different frames for classical and quantum particles, as well as spin and spinless particles, such as bosons and fermions. \newline

Gravitational effects on quantum fields have been rigorously studied for a few decades, particularly in the case of spinor fields. Standard books such as \cite{Birrell:1982,penrose:spinor,Wald:1995yp,Diraceq} delve into the mathematical structure of the spinor formalism. Spinor fields in curved space-times have been studied in several papers, and we make a brief review of these works. In \cite{RevModPhys.25.714} the authors develop the formalism of the Dirac equation in a curved space-time coupled to an electromagnetic field.  In \cite{deOliveira:1962apw} the authors  give the key to generalize the Dirac equation from flat space-time to general relativity via the tetrad formalism with the Lorentz invariant transformation. 
In \cite{PhysRevD.23.2157,PhysRevD.22.1922} the authors study the quantum mechanics of the hydrogen atom in a general relativistic context. In \cite{PhysRevD.23.2157} the analog of the Stark effect is considered with the  center of mass formalism. Paper \cite{PhysRevD.22.1922} analyses the modifications in the eigenvalues of the energy spectrum that arise due to the curvature of space-time. Additionally,\cite{RevModPhys.29.465} compares the energy levels of neutrinos and electrons in a curved space-time with spherical symmetry, that is, the Schwarzschild metric. Moreover, the authors study  thermodynamical processes  and the creation of neutrino pairs. On the other hand, in paper \cite{10.1143/PTP.100.1145} the authors write the Dirac and Weyl equations for neutrinos in a Kerr metric using the tetrad formalism and compare them with the results obtained in a spherical metric without rotation. We mentioned  these references to place our work in a broader context. There are specific points that we shall discuss deeply in the next sections, one of them being the consistency conditions for the continuity equation. More information about the continuity equation can be found  in \cite{Arminjon1,Arminjon2,Arminjon3,Pollock}.\newline

This paper is organized as follows. In section \ref{section: Field equations}, we present the field equations and the formalism that we will use to describe the Dirac fermions in curved space-times. In section \ref{section: Dirac Hydrodynamic representation}, we introduce a generalized Madelung transformation for Dirac fermions, which implies a hydrodynamic representation for this case. Since, we can work using either the Dirac or Weyl representation for 1/2-spin fermions. In section \ref{section: Weyl representation}, we give a brief introduction to these both representations, further it is shown the field equations for the Weyl fermions (or the chiral form of the Dirac fermions). Analogously, for Weyl fermions we introduce the hydrodynamic representation from a generalized Madelung transformation in section \ref{section:Weyl Hydrodynamic representation}. For both kinds of fermions in section \ref{section: energy balance}, we explain what are the different contributions of the energy for a Fermi gas in a curved space-time coupled to an electromagnetic field and we show a generalized Gross-Piitaevskii equation for fermions. Moreover, the conclusions are indicated in section \ref{section: conlusions} and the acknowledgments are shown in section \ref{section:Acknowledgments}. Finally, in appendix \ref{appendixA}, we can find a solution for a simple example to the Dirac equation in a flat space-time. 

\section{Field Equations}
\label{section: Field equations}

We start using the tetrad formalism for the space-time geometry, and the canonical expansion of the space-time in a 3+1 ADM decomposition \cite{Diraceq,ADM_formalism,Alcubierre,HawkingEllis:1973,Nakahara:2003,Gallegos:2019pyf}, such that the coordinate $t$ is the parameter of evolution. The 3+1 metric reads
\begin{equation}
\label{metric}
	\mathrm{d}s^2 =  N^2 c^2 
	\mathrm{d}t^2 - h_{ij} \left(\mathrm{d}x^i + N^i c \, \mathrm{d}t \right) \left(
	\mathrm{d}x^j + N^j c \, \mathrm{d}t \right) ,
\end{equation}
where $N$ represents the lapse function which measures the proper time of the observers traveling along the world line, $N^i$ is the shift vector that measures the displacement of the observers between the spatial slices and $h_{ij}$ is the 3-dimensional slice-metric. In what follows $i,j,k,l=1,2,3$ are the spatial indices; $a,b,c=0,1,2,3$ and $\mu,\nu,\alpha=0,1,2,3$ the space-time indices. We write eq. (\ref{metric}) in the tetrad formalism as d$s^2=\eta_{ab}e^a_{\,\,\mu} e^b_{\,\,\nu} dx^\mu dx^\nu$, where $\eta_{ab}=$ diag$(1,-1,-1,-1)$. Here  $e^a=e^a_{\,\,\mu} dx^\mu$ is the set of one-forms base of the cotangent space at the space-time manifold given by
\begin{eqnarray}
\label{one-form}
	e^0 &=& N c\mathrm{d}t, \nonumber \\
	e^k &=& \hat e^k_{\,\,\,i}\left(\mathrm{d}x^i + N^i c \, \mathrm{d}t \right),
\end{eqnarray}
with inverse
\begin{eqnarray}
\label{one-formInverse}
	e_0 &=& \frac{1}{N} \left(\frac{\partial}{c\, \partial t}- N^j \frac{\partial}{\partial x^j}\right),\nonumber \\
	e_k &=&  \hat{e}_k^{\,\,\,j}\frac{\partial}{\partial x^j},
\end{eqnarray}
where $\hat e^k=\hat e^k_{\,\,i}\mathrm{d}x^i$ are the one-form base to the three-dimensional slice of the cotangent manifold, such that $h_{ij}=\delta_{kl}\hat e^k_{\,\,i} \hat e^l_{\,\,j}$. We can also define the set of vectors base of the tangent-space to the space-time as $e_a=e^{\,\,\mu}_a\partial_\mu$, such that $e^a e_b=\delta^a_{\,\,b}$. We will use the tetrad formalism\cite{penrose:spinor,Diraceq,HawkingEllis:1973,Nakahara:2003,Gallegos:2019pyf,Wald:1984} to describe the space-time geometry where the fermion particles are located.\newline

The action of a fermion system  in curved space-times coupled to an electromagnetic field $A_\mu$ is given by $S\left[\psi(x^\mu),\partial_\mu\psi(x^\mu)\right]=\int\mathcal{L}\left(\psi(x^\mu),\partial_\mu\psi(x^\mu)\right)d^4x$, where $\mathcal{L}=\mathcal{L}\left(\psi(x^\mu),\partial_\mu\psi(x^\mu)\right)$ is the Lagrangian density   \cite{Arminjon1,Arminjon2,Arminjon3}:
\begin{align}
    \label{Dirac lagrangian} 
   \mathcal{L}=\sqrt{-g}\dfrac{i\hbar c}{2}\left[\psi^\dagger B\gamma^\mu\left(D_\mu\psi\right)-\left(D_\mu\psi\right)^\dagger B\gamma^\mu\psi+\dfrac{2imc}{\hbar}\psi^\dagger B\psi\right].
\end{align}
Here, $D_\mu=\nabla_\mu+\dfrac{iq}{\hbar c}A_\mu$ is the total covariant derivative accounting for electromagnetic effects. The covariant derivative of a spinor $\psi=(\psi_{\dot\nu})$ is given by $\nabla_\mu (\psi_{\dot\nu})=\partial_\mu (\psi_{\dot\nu})+\Gamma^{\dot\alpha}_{\mu\dot\nu}(\psi_{\dot\alpha})$, where $\Gamma^{\dot\alpha}_{\mu\dot\nu}$ is the spin connection\cite{Diraceq,Poplawski:2007iz}. Observe the internal indices as dot indices. Using the least action principle it is possible to obtain from eq.(\ref{Dirac lagrangian}) the corresponding Dirac equation. This equation is given by
\begin{equation}
\label{Dirac eq}
    \left[i\hbar\gamma^\mu\right(\nabla_\mu+iqA_\mu\left)-mc\right]\psi=0,
\end{equation}
where $\hbar$, $c$ are the Planck constant and the speed of light respectively, while $q, m$ are the charge and mass of the fermion particle and $\psi$ is its spinor. Besides, the gamma matrices $\gamma^\mu$ are related to the spin and space-time geometry. They can be written as $\gamma^\mu=e^\mu_{\,\,a}\Tilde{\gamma}^a$, where $\Tilde{\gamma}^a$ are the gamma matrices in flat space-time, which are well-know from standard Quantum Field Theory (QFT) \cite{Weinberg:1995,aitchison,Lancaster:2014} Henceforth, to simplify the notation, we use the natural units ($c=\hbar=1$), instance, $mc/\hbar\rightarrow m$. Therefore,
\begin{eqnarray}\label{eq:defgammas}
    \gamma^0&=&N\Tilde{\gamma}^0, \nonumber  \\
    \gamma^k&=&\hat{e}^k_{\,\,\,j}(\Tilde{\gamma}^j+N^j\Tilde{\gamma}^0). 
\end{eqnarray}
In general, these matrices fulfill the following anti-commutation relation\cite{Birrell:1982}\cite{Diraceq}
\begin{equation}\label{eq:gammas=2g}
    \{\gamma^\mu,\gamma^\nu\}=\gamma^\mu\gamma^\nu+\gamma^\nu\gamma^\mu=2g^{\mu\nu}\mathbb{I},
\end{equation}
where $g_{\mu\nu}$ represents the metric that describes the space-time geometry. Furthermore, as we know, the gamma matrices in flat space-time are related to the Pauli matrices, which describe the spin of the fermion particles. In addition, due to the Lorentz invariance that spinors follow, we note that $\psi\psi^\dagger$ is not a Lorentz scalar and neither $\psi\gamma^\mu\psi^\dagger$ is a Hermitian. On the other hand, we observe that, in general, the gamma matrices obey the following relation\cite{Arminjon1,Arminjon2,Arminjon3,Pollock} 
\begin{equation}
\label{hermitizing relation}
    (\gamma^\mu)^\dagger=B\gamma^\mu B^{-1},
\end{equation}
where $B$ is a hermitian matrix, i.e. $B^\dagger=B$, that is uniquely determined by the gamma matrices $\gamma^\mu$. As usual, we denote by $B^\dagger$ the conjugate (or Hermitian) transpose of $B$. In contrast, using eq.(\ref{hermitizing relation}) it is straightforward to obverse the invariant quantities under the Lorentz transformation are $\psi\bar{\psi}$ as scalar and $\psi\gamma^\mu\bar{\psi}$ as a four-vector, where $\bar{\psi}=\psi^\dagger B$ is named the adjoint spinor (see more in\cite{Birrell:1982,penrose:spinor,Diraceq,Weinberg:1995}).\newline

Furthermore, we note that in QFT the relation (\ref{hermitizing relation}) is fulfilled when $B=\Tilde{\gamma}^0$ and the gamma matrices are in flat space-time. From the action (\ref{Dirac lagrangian}) of the fermion system we can find the equation for the transpose conjugated spinor by making an infinitesimal variation of this action with respect to $\psi$. Another way of getting this equation of motion is to take the transpose conjugate of the Dirac equation (\ref{Dirac eq}) and using (\ref{hermitizing relation}). In this manner we find that the transpose conjugated Dirac equation in curved space-time is given by
\begin{equation}
    \label{transpose Dirac eq}
    i\left(\nabla_\mu\bar{\psi}\right)\gamma^\mu-i\psi^\dagger\nabla_\mu\left(B\gamma^\mu\right)+i\bar{\psi}\nabla_\mu\gamma^\mu+\bar{\psi}A_\mu\gamma^\mu+m\bar{\psi}=0.
\end{equation}
We consider $(\nabla_\mu\psi)^\dagger=\nabla_\mu\psi^\dagger$ and denote the adjoint spinor as  $\bar{\psi}=\psi^\dagger B$. Using the gamma matrices in flat space-time and the fact that $B=\Tilde{\gamma}^0$, we recover the  definition of $\bar{\psi}$ in QFT and the transpose conjugated Dirac equation. However, in an arbitrary space-time $\nabla_\mu\gamma^\mu$ is distinct from zero, since $\gamma^\mu=e^\mu_{\,\,a}\Tilde{\gamma}^a$. Therefore, in general $\nabla_\mu e^\mu_{\,\,a}$ is non-zero.\newline

We can get the conserved charge from the Noether theorem  \cite{Noether}. The Dirac current is
\begin{equation}
    \label{dirac current}   
    J^{\mu}=\bar{\psi}\gamma^\mu\psi=\psi^\dagger B\gamma^\mu\psi.
\end{equation}
To obtain the continuity equation 
\begin{equation}
\label{conteq}
\nabla_\mu J^{\mu}=0,
\end{equation}
for the Dirac current, we take the covariant derivative of eq. (\ref{dirac current}). This gives
\begin{equation}
\label{continuity eq}
    \nabla_\mu J^{\mu}=(\nabla_\mu\bar{\psi})\gamma^\mu\psi+\bar{\psi}\left(\nabla_\mu\gamma^\mu\right)\psi+\bar{\psi}\gamma^\mu\nabla_\mu\psi.
\end{equation}{}
If we multiply the Dirac equation  (\ref{Dirac eq}) by $\bar{\psi}$ and its transpose conjugate (\ref{transpose Dirac eq}) by $\psi$ and sum both equations, it follows that 
\begin{equation}\label{eq:conservacion}
    \nabla_\mu J^{\mu}=\psi^\dagger\nabla_\mu\left(B\gamma^\mu\right)\psi.
\end{equation}
If we require that the continuity equation (\ref{conteq})
is fulfilled, i.e., that the number of particles is conserved, then we need $\nabla_\mu\left(B\gamma^\mu\right)=0$, or equivalently
\begin{equation}
\label{B-gamma derivative eq}
    (\nabla_\mu B)\gamma^\mu=-B\nabla_\mu\gamma^\mu.
\end{equation}
At this point, we want to emphasize the consistency conditions for the continuity equation (\ref{conteq}). Some authors in \cite{RevModPhys.29.465} impose $\nabla_\mu \gamma^\nu=0$ while others, \cite{PhysRevD.22.1922}, impose $\nabla_\mu B=0$. These conditions are independent of each other. Instead, in references \cite{Arminjon2,Arminjon3}, the authors conclude that the condition $\nabla_\mu(B\gamma^\nu)=0$ is the most convenient because it is implied by $\nabla_\mu\gamma^\nu=0$ and  $\nabla_\mu B=0$. \newline

In addition, we can note that the matrix $B$ can be obtained for a general metric (\ref{metric}) by solving  the differential equation
\begin{equation}
\label{B gamma descomposition derivative eq}
    \left(\nabla_0(BN)+\nabla_j(B\hat{e}^j_iN^i)\right)\Tilde{\gamma}^0-\nabla_j(B\hat{e}^j_i)\Tilde{\gamma}^i=0,
\end{equation}
which follows from eq. (\ref{B-gamma derivative eq}). Using the condition (\ref{B-gamma derivative eq}), it is possible to rewrite the transpose conjugated Dirac equation (\ref{transpose Dirac eq}) as
\begin{equation}
    \label{transpose conjugated Dirac eq}
    i\left(\nabla_\mu\bar{\psi}\right)\gamma^\mu+i\bar{\psi}\nabla_\mu\gamma^\mu+\bar{\psi}A_\mu\gamma^\mu+m\bar{\psi}=0.
\end{equation}
In order to find the conserved quantity resulting from the continuity equation, we take an arbitrary surface $\mathcal{S}$ enclosing the volume $\mathcal{V}$ which contains the whole system. Let $k^j$ be an orthonormal vector to $\mathcal{S}$ such that 
\begin{equation}
    \label{integral continuity eq}
    \int_\mathcal{V}\nabla_\mu J^{\mu} dV=\int_\mathcal{V}\nabla_0 J^{0} dV+\int_\mathcal{S}k_j J^{j}\sqrt{h} d^3x=0.
\end{equation}
where $h$ is the determinant of the slice-metric $h_{ij}$. We assume that far away from the source spinor $\psi$ goes to zero, that means that in this region $J^{\mu}$ is negligible. Then, the surface integral in eq. (\ref{integral continuity eq}) vanishes, and we obtain
\begin{equation}
    \label{conserved charge}
    \dfrac{dQ}{dt}=\int_\mathcal{V}\nabla_0 J^{0} dV=0,
\end{equation}
where $Q=\int_\mathcal{V} J^{0} dV$ is the conserved charge, $dV$ is the curved volumen element $dV=\sqrt{-g}d^4x$. In QFT this charge is identified with the number of fermions or with the electric charge of the system. In flat space-time we have  $B=\Tilde\gamma^0$, so that $J^0=\psi^\dagger \psi=n$ represents the number density of fermion particles. In curved space-time $J^0$ (which is  determined by  $\gamma^0$ and by the generalized gamma matrices) has a different interpretation. The form of $B$  given by eqs. (\ref{hermitizing relation}) and (\ref{B-gamma derivative eq}) for each metric is related to the gamma matrices and to the tetrad formalism.  \newline

Finally, since the spinor field used is coupled to an electromagnetic field, we show the equations that describe the electromagnetic field. Thus, with the Maxwell four-potential we can define the Faraday tensor
\begin{equation}
\label{faraday tensor}
    F_{\mu\nu}=\nabla_\mu A_\nu-\nabla_\nu A_\mu.
\end{equation}
In the electromagnetic theory, the Faraday tensor $F_{\mu\nu}$ satisfies the Maxwell field equations
\begin{equation}
\label{Maxwell eq}
    \nabla_\nu F^{\nu\mu}=J^{E\mu},
\end{equation}
where $J^{E\mu}$ is the four-electromagnetic current.\newline

At this point, we gave the most general form for standard Dirac fermions in an arbitrary framework coupled to an electromagnetic field. In fact, for quantities like $B$ and $\gamma^\mu$ we have not yet adopted any representation. Nevertheless, we will have to make this decision to give some examples and results in the sections below.

%The problem of the Energy Balance for boson particles in a curved space-time is studied in \cite{Matos:2016ryp}, where the conserved 4-current associated with the KG equation describing the evolution of a complex scalar field $\Phi(x^\mu)$ is defined. We can generalize this idea by defining a new 4-current $J^{KG}_\mu$, changing the scalar field by a spinor and the complex conjugate scalar field by the conjugate transpose of the spinor. Namely, the KG current is redefined as
%\begin{equation}
%    \label{KG current}
%    J^{KG}_\mu=i\dfrac{q}{2m^2}\left[\psi\left(\nabla_\mu-iqA_\mu\right)\psi^\dagger-\psi^\dagger\left(\nabla_\mu+iqA_\mu\right)\psi\right].
%\end{equation}

\section{Dirac Hydrodynamic Representation}
\label{section: Dirac Hydrodynamic representation}

Analogously to the hydrodynamic representation of the Schr\"odinger equation, which was introduced by Madelung\cite{Madelung}, we derive the hydrodynamic representation of the Dirac equation. We carry out the following generalized Madelung transformation for each component of the spinor $\psi=\psi(x^\mu)$ as follows
\begin{equation}
    \label{Madelung transformation}
    \psi=\exp(i\theta\mathbb{I})R,
\end{equation}
where $\mathbb{I}$ is the identity matrix, $R$ is a spinor and $\theta$ is a complex function. Observe that the spinor $\psi$ has eight degrees of freedom and the spinor $R\exp(i\theta\mathbb{I})$ has ten. A similar situation appeared for the case of the boson case, where the scalar field $\Phi=\Psi\exp(i\theta)$ has two degrees of freedom and the right hand side has three. This extra degree of freedom is interpreted as the velocity potential. Here it will be a similar situation. 
In what follows we will denote $\theta\mathbb{I}\rightarrow \theta$, unless it is specify. For the case where we consider a Dirac electron-like fermion, $\theta=\theta(x^\mu)$, the spinor $\psi$ reads
\begin{equation}
  \label{Madelung transformation1}
    \psi= \left( \begin{array}{cccc} 
    R_{\dot{1}} \\ 
    R_{\dot{2}} \\ 
    R_{\dot{3}} \\ 
    R_{\dot{4}} 
    \end{array} \right)\exp(i\theta)=R\exp(i\theta),
\end{equation}

where we use the notation $\dot\mu$, $\dot\nu$, ...$=\dot 1,\cdots,\dot 4$ for the spinor indices such that
\begin{equation}
  \label{eq:Rdefinicion}
    R= \left( \begin{array}{cccc} 
    R_{\dot{1}} \\ 
    R_{\dot{2}} \\ 
    R_{\dot{3}} \\ 
    R_{\dot{4}} 
    \end{array} \right)=\left(
    \begin{matrix}
    \sqrt{n_{\dot{1}} }\\ 
    \sqrt{n_{\dot{2}}} \\
    \sqrt{n_{\dot{3}}} \\
    \sqrt{n_{\dot{4}}} 
    \end{matrix}
    \right).
\end{equation}
where we will use $n_{\dot{\mu}}=|R_{\dot{\mu}}|^2$,
%\textcolor{blue}{Here, the indices do not use the sum convention when two or more quantities together labeled with the same index, where both are subindices or upper indices, in this case, only they indicate a label. Otherwise, the sum convention follows.} 
here $n_{\dot\mu}$ is the number density which represents the modulus of $\psi_{\dot\mu}$ and $\theta$ is its phase (both are complex variables). In general, $n_{\dot\mu}$ is different for each component of the spinor. Note that the covariant derivative of the spinor $\psi$ in terms of its decomposition (\ref{Madelung transformation1}) is $\nabla_\mu (\psi_{\dot\nu})=\partial_\mu (R_{\dot\nu}e^{i\theta})+\Gamma^{\dot\alpha}_{\mu\dot\nu}(R_{\dot\alpha}e^{i\theta})=(\partial_\mu R_{\dot\nu})e^{i\theta}+ i(\partial_\mu\theta)R_{\dot\nu}e^{i\theta}+\Gamma^{\dot\alpha}_{\mu\dot\nu}(R_{\dot\alpha}e^{i\theta})$, implying that $\nabla_\mu \theta=\partial_\mu \theta$. In the appendix, we show some exact solutions of the Dirac equation with this ansatz in flat space-time.\newline

Using the transformation (\ref{Madelung transformation1}) in eq. (\ref{Dirac eq}), the Dirac equation in terms of the variables $R$ and $\theta$ reads
\begin{eqnarray}
\label{hydro dirac eq_b}
     \exp(i\theta)\gamma^\mu\left(i{\nabla}_\mu R-(\nabla_{\mu}\theta)R-q{A}_\mu R-\frac{m}{4}\gamma_\mu R\right)=0.
    %\\    i\slashed{\nabla}R-\gamma^{\mu}R\nabla_{\mu}\theta-q\slashed{A}R-mR&=&0,
\end{eqnarray}
To get the last term, we used the property of the gamma matrices that $\gamma_\mu\gamma^\mu=4\mathbb{I}$, where $\mathbb{I}$ is the $4\times 4$ identity matrix. This property results from the anti-commutation relation of the gamma matrices.  

Similarly,   the continuity equation (\ref{conteq})
 with (\ref{dirac current})  can be written with these new variables as
\begin{eqnarray}
\label{hydro dirac eq1}
\left(\nabla_{\mu} R^\dagger\right)K^{\mu} R+R^\dagger K^{\mu}\left(\nabla_{\mu} R\right)=0,
\end{eqnarray}
where $R^\dagger$ denotes the conjugated transpose of $R$ and $K^{\mu}=B\gamma^{\mu}$. Observe that $K^\mu$ is hermitian ($K^{\mu\dag}=K^\mu$). 
%In addition, $K^\mu=B\gamma^{\mu}$ since the matrix $\exp(i\theta)$ is diagonal.\newline 

%Finally, \textcolor{red}{we can re-write equation (\ref{hydro dirac eq1}) in components,} since  the spinor and conjugate spinor are different from zero,  the Dirac equation and the transposed Dirac equation take the form 
%\begin{align}
   % \label{Madelung Dirac eq}
   % \dfrac{i}{2}\slashed{\nabla}\ln{(n_{\dot\nu})}-\slashed{\nabla}\theta-q\slashed{A}-m=0,\\
  %  \label{Madelung trans Dirac} 
  % \dfrac{i}{2}\nabla_\mu\ln{(n^{\dot\nu})}\gamma^{\mu}+\nabla_\mu\theta\gamma^{\mu}+q A_\mu\gamma^{\mu}+m=0,
%\end{align}
%where we introduced the notations $\slashed{\nabla}=\gamma^\mu\nabla_\mu$ and $\slashed{B}=\gamma^{\mu}B_{\mu}$ for any vector $B_\mu$.  Although the values for each component of the spinor and its transpose are equal, namely $R_{\dot\alpha}=R^{\dot\alpha}$, 
%and the same for $\theta$, which is a real diagonal matrix $\theta=\theta^T$,we use the notation  $\left(R^T\right)_{\dot\alpha}= R^{\dot\alpha}$ and $ (n^T)_{\dot{\alpha}}=n^{\dot\alpha}$ to make a distinction between the internal components of $R$ and its transpose $R^T$, the same for $n$ and $n^T$. \newline

Summarizing, we have introduced the Madelung transformation for the Dirac equation (\ref{hydro dirac eq_b}) and the continuity relation (\ref{hydro dirac eq1})
%(\ref{Madelung trans Dirac}) 
by making the change of variables from eq. (\ref{Madelung transformation}). With this new form to write the Dirac equation, we can introduce variables that have a more plausible physical interpretation in quantum theory.  \newline

To see this, we apply the operator $i\gamma^\mu D_\mu=i\gamma^\mu\nabla_\mu-q\gamma^\mu A_\mu$ to the Dirac equation (\ref{Dirac eq}) written under the form $i\gamma^\mu\nabla_\mu\psi=q\gamma^\mu A_\mu\psi+m\psi$. This yields
\begin{eqnarray}\label{eq:diracKG1}
-\gamma^\mu\gamma^\nu\left(\nabla_\mu\nabla_\nu\psi+iq(\nabla_\mu A_\nu)\psi+iqA_\nu(\nabla_\mu\psi)+iqA_\mu(\nabla_\nu\psi)-q^2A_\mu A_\nu\psi\right)&-&\nonumber\\
       m^2\psi-\gamma^\mu(\nabla_\mu\gamma^\nu)(\nabla_\nu\psi+iqA_\nu\psi)&=&0.\nonumber\\
\end{eqnarray}
Using the relation (\ref{eq:gammas=2g}) in eq. (\ref{eq:diracKG1}), we obtain
\begin{eqnarray}
      \label{eq:diracKG2}
      \Box_E\psi+m^2\psi+
      \frac{i}{2}q\gamma^\mu\gamma^\nu F_{\mu\nu}\psi+\gamma^\mu(\nabla_\mu\gamma^\nu)(D_\nu\psi)=0,
\end{eqnarray}
where we have defined the D'Alambertian operator in the presence of an electromagnetic field by $\Box_E=(\nabla_\mu+iq A_\mu)(\nabla^\mu+iq A^\mu)$ and the anti-symmetric Faraday tensor by $F_{\mu\nu}=\nabla_\mu A_\nu-\nabla_\nu A_\mu$. Eq. (\ref{eq:diracKG2}) is similar to the Klein-Gordon equation with an electromagnetic source except that here $\psi$ is a spinor instead of a complex scalar field. Note that the first two terms in (\ref{eq:diracKG2}) are the Klein-Gordon equation, but the the electromagnetic field and the spinorial character of the equation add two more terms. The difference here is that if you ``square" the Dirac equation in flat space-time, you obtain the Klein-Gordon equation, for an arbitrary curved space this does not happen. The last term of eq.  (\ref{eq:diracKG2}) contains the covariant derivative of $\gamma^\mu$ which vanishes in a flat space-time.\newline

As for the Klein-Gordon equation \cite{Matos:2016ryp,Chavanis:2016shp}, we define the diagonal matrix 4-velocity $v_\mu$ by
\begin{equation}
    \label{4-velocity}
    mv_{\mu}=\nabla_\mu S+qA_\mu\mathbb{I}.
\end{equation}
Here, $S(x^\mu)$ is a phase with components $S=(\theta-\omega t)\mathbb{I}$, where $\omega$ are constants that can be related to the mass of the fermion particle by $\omega=mc^2/\hbar$. In this manner we can write
\begin{equation}
    \label{4-velocity2}
    \nabla_\mu \theta\mathbb{I}=mv_{\mu}-\omega\delta^{0}_{\,\,\,\mu}\mathbb{I}-qA_\mu\mathbb{I}.
\end{equation}
%With these new variables, the Dirac equations (\ref{Madelung Dirac eq}) and (\ref{Madelung trans Dirac}) become
%\begin{align}
   % \label{new dirac eq}
   %\dfrac{i}{2}\slashed{\nabla}\ln{(n_{\dot\nu})}-  m\slashed{v}_{\dot\nu}-\omega_{\dot\nu}\slashed{\nabla} t-m=0,\\
   %\label{new transp dirac eq}
   %\left(\dfrac{i}{2}\nabla_\mu\ln{(n^{\dot\nu})}+ (mv^{\,\,\,\dot\nu}_{\mu}+\omega^{\dot\nu}\nabla_\mu t)+\dfrac{m}{4}\gamma_{\mu}\right)\gamma^\mu=0.
%\end{align}
%Equation (\ref{new dirac eq}) is the Dirac equation in the new variables $n_{\dot{\nu}}$, $v_{\mu\dot{\nu}}$, where 
We interpret $n_{\dot{\nu}}$ as the density number of fermions and $v_{\mu}$ as its velocity. In what follow we denote $\omega\rightarrow\omega\mathbb{I}$ unless otherwise stated. Additionally, we will show that eq.(\ref{hydro dirac eq_b}) can be interpreted as the first integral of the Bernoulli equation for fermions in an arbitrary space-time. For doing so, we will use this new interpretation using variables $n_{\dot{\nu}}$ and $v_\mu$ in the Dirac equation, instead of $\psi$ in order to write  a Navier-Sotkes-like equation for ferminos, in the same way a it was done for bosons in \cite{Matos:2016ryp}. Then, we will see that equation (\ref{hydro dirac eq1}) can be interpreted as the generalized first integral of the Bernoulli equation in the sense that, for obtaining the Navier-Stokes-like equation, we need to differentiate equation (\ref{hydro dirac eq_b}).\newline

According to \cite{Matos:2016ryp,Chavanis:2016shp} if we apply the transformation (\ref{Madelung transformation}) to eq. (\ref{eq:diracKG2}), we could expect to obtain the continuity equation for the imaginary part  and the Bernoulli equation for the real part. However, in the case of the Dirac equation, the four components are mixed by the presence of the four dimensional spinor $\psi$. Hence, we obtain the following expression
\begin{eqnarray}\label{eq:diracKG3}
     i\left[ 2(m v^{\mu}-\omega\delta^\mu_0)\nabla_\mu R-qA_\mu+q\nabla_\mu(A^\mu R)+\nabla_\mu(mv^\mu-\omega\delta_0^\mu-qA^\mu)R\right]&+&\nonumber\\
     \left(m^2v_{\mu}{v^{\mu}}+2m\omega{v^0}+\frac{\omega^2}{N^2}+m^2\right)R-\Box R&+&\nonumber\\
     \frac{i}{2}q\gamma^\mu\gamma^\nu F_{\mu\nu} R+\gamma^\mu(\nabla_\mu\gamma^\nu)(i(mv_{\nu}+\omega \nabla_\nu t) R+D_\nu R)&=&0.\nonumber\\
\end{eqnarray}
Here, we have defined $\Box=\nabla^\nu\nabla_\nu$.
%and we have introduced the diagonal matrices $v_\mu=v_{\mu\dot{\nu}}\delta^{\dot{\nu}}_{\dot\alpha}$ and $\omega=\omega_{\dot{\nu}}\delta^{\dot{\nu}}_{\dot\alpha}$. 
For bosons, the real and imaginary parts are separated into two independent equations, namely, the continuity equation and the Bernoulli equation \cite{Matos:2016ryp,Chavanis:2016shp}. But in the spinor case, the last line of equation (\ref{eq:diracKG3}) mixes both the imaginary and real parts and there is no natural separation into real and imaginary parts. The system remains coupled.\newline

\section{Weyl Representation}
\label{section: Weyl representation}
The Dirac equation for $1/2$-spin particles is associated with the $SO(1,3)$ symmetry group. Nevertheless, we can introduce a new representation as in standard QFT, since there exists a surjective homomorphism between the $SO(1,3)$ and $SU(2)\otimes SU(2)$ Lie groups.\newline 

As we know, the special unitary group $SU(2)$ is formed by the set of $2\times 2$ complex matrices $A$, which satisfy det$(A)=1$. Explicitly, we have
\begin{equation}
A=\left(
\begin{matrix}
a & -\bar{b}\\
b & \bar{a}
\end{matrix}
\right),
\end{equation}
with det$(A)=|a|^2+|b|^2=1$, where $a$ and $b$ are complex parameters. Equivalently, we have the identity $A^\dagger=A^{-1}$.\newline 

The Lie algebra $\mathfrak{su}(2)$ associated to the $SU(2)$ Lie group is given by the exponential map
\begin{equation}
    \label{su(2) algebra}
    \exp(\mathfrak{su}(2))\rightarrow SU(2).
\end{equation}
For any element $X$ of the Lie algebra, we have $\exp(X)\exp(X)^\dagger=\mathbb{I}$, implying that $X+X^\dagger=0$. In what follows, we will indistinctly use $\exp(X)$ and $e^X$ as the exponential map.\newline

In terms of the Pauli matrices $\sigma^\mu$ the $4\times 4$ gamma matrices $\gamma^\mu$ can be written as two $2\times 2$ block matrices 
\begin{eqnarray}
\label{Weyl gamma matrix}
\gamma^0&=&N\Tilde{\gamma}^0=
N\left(\begin{matrix}
    0 & \mathbb{I}\\ 
    \mathbb{I} & 0
    \end{matrix}\right),\\
\gamma^j&=&\hat{e}^j_{\,\,\,i}(\Tilde{\gamma}^i+N^i\Tilde{\gamma}^0)=\left(
    \begin{matrix}
    0 & -\hat{e}^j_{\,\,\,i}(\Tilde{\sigma}^i-N^i\mathbb{I})\\ 
    \hat{e}^j_{\,\,\,i}(\Tilde{\sigma}^i+N^i\mathbb{I}) & 0
    \end{matrix}
    \right),
    \label{Weyl gamma matrix2}
\end{eqnarray}
where $\Tilde{\sigma}^i$ are the $2\times 2$ Pauli matrices in flat space-time 
\begin{equation}\label{eq:Pauli}
     \Tilde{\sigma}^1=\left(
    \begin{matrix}
    0 & 1\\ 
    1 & 0
    \end{matrix}
    \right),\,\,\,\, 
    \Tilde{\sigma}^2=\left(
    \begin{matrix}
    0 & -i\\ 
    i & 0
    \end{matrix}
    \right),\,\,\,\,
     \Tilde{\sigma}^3=\left(
    \begin{matrix}
    1 & 0\\ 
    0 & -1
    \end{matrix}
    \right),
\end{equation}
and $\mathbb{I}$ is the $2\times 2$ identity matrix. The $\gamma^\mu$ matrices satisfy $\left(\gamma^0\right)^\dagger=\gamma^0$ and $\left(\gamma^{j}\right)^\dagger=-\gamma^j+2N^j\gamma^0/N$. At this point, we need to adopt the standard representation for the gamma matrices in a flat space-time $\Tilde{\gamma}^\mu$ as follows  
\begin{equation}\label{eq:gamma in flat space}
     \Tilde{\gamma}^0=\left(
    \begin{matrix}
    0 & \mathbb{I}\\ 
    \mathbb{I} & 0
    \end{matrix}
    \right),\,\,\,\, 
    \Tilde{\gamma}^j=\left(
    \begin{matrix}
    0 & -\Tilde{\sigma}^j\\ 
    \Tilde{\sigma}^j & 0
    \end{matrix}
    \right).
\end{equation}
This representation helps us to build the Weyl representation. Additionally, in the Weyl representation we can write a Dirac fermion as a four-spinor $\psi$ made of two spinors, each of which having  two components, for instance
\begin{equation}
    \label{Weyl spinor}
    \psi= \left( \begin{array}{cc} \psi_R \\ \psi_L \end{array} \right),
\end{equation}
where $\psi_R$ and $\psi_L$ are the right- and the left- handed Weyl spinors, respectively. If we write the adjoint spinor $\bar{\psi}$ and use the Weyl representation, it follows that
\begin{equation}
    \label{Weyl general adjoint spinor}
    \bar{\psi}=\psi^\dagger B= \left(  \psi_R^\dagger , \psi_L^\dagger  \right)B,
\end{equation}
where $B$ is the matrix from eqs.  (\ref{hermitizing relation}) and (\ref{B-gamma derivative eq}). If we use the relation (\ref{hermitizing relation}) it is straightforward to see that the matrix $B$ must have the following form
\begin{equation}
    \label{eq:Bforma}
    B= \left( \begin{array}{cc} 0 & B_\zeta \\ 
    B_\zeta & 0 \end{array} \right),
\end{equation}
where the $2\times 2$ matrix $B_\zeta$ is a diagonal matrix, $B_\zeta=b\mathbb{I}$, with $b=b(x^\mu)$. Therefore, we get $B=b\Tilde{\gamma}^0$ and eq. (\ref{B gamma descomposition derivative eq}) transforms into
\begin{eqnarray}
    \label{eq:Bzeqdif}
    \nabla_0(Nb)+\nabla_j(\hat{e}^j_iN^ib)&=&0,\\
    \nabla_j(\hat{e}^j_ib)\Tilde{\sigma}^i&=&0.
\end{eqnarray}
Note that in eq.(\ref{eq:Bzeqdif}), we assume also a representation to $B$ matrix. Adopt a specific representation for the symmetry group, which is done without loss of generality. In fact, it shall make this choice to build the Weyl fermions and its field equations. Hence, using the definition of the spinor and its adjoint we can write the Dirac quadricurrent  $J^{\mu}$ from eq. (\ref{dirac current}) as
\begin{equation}
    \label{Weyl general current}
    J^\mu=\left(\psi_R^\dagger, \psi_L^\dagger\right)B\gamma^\mu \left(\begin{array}{cc}
         \psi_R\\
         \psi_L 
    \end{array}\right),
\end{equation}
where the gamma matrices are defined by eqs. (\ref{Weyl gamma matrix}) and (\ref{Weyl gamma matrix2}) and, in general, $B$ is given by the previously mentioned conditions. This yields
\begin{eqnarray}
    \label{eq:Dirac-current}
    J^0&=&Nb(\psi^\dagger_R \psi_R+\psi^\dagger_L \psi_L),\\
    J^j&=&b\hat{e}^j_{\,\,\,i}(\psi^\dagger_R (\Tilde{\sigma}^i+N^i\mathbb{I}) \psi_R-\psi^\dagger_L (\Tilde{\sigma}^i-N^i\mathbb{I}) \psi_L).
\end{eqnarray}
In order to simplify the notation, we now define the vectors of $2\times2$ matrices $\mathbb{S}^a=(\mathbb{I},\Tilde{\sigma}^j+N^j\mathbb{I})$ and  $\mathbb{\bar{S}}^a=(-\mathbb{I},\Tilde{\sigma}^j-N^j\mathbb{I})$ in terms of the Pauli matrices. $\mathbb{S}^a$ and $\mathbb{\bar{S}}^a$ are the (generalized) Pauli matrices in flat space-time. In terms of these new definitions, the density currents read 
\begin{eqnarray}
    \label{eq:Dirac-current1}
    J^\mu &=&b\hat{e}^\mu_{\,\,\,i}(\psi^\dagger_R \mathbb{S}^i \psi_R-\psi^\dagger_L \bar{\mathbb{S}}^i \psi_L)\nonumber\\
    &=&b(\psi^\dagger_R \sigma^\mu \psi_R-\psi^\dagger_L \bar{\sigma}^\mu \psi_L),
\end{eqnarray}
where we have defined the $2\times 2$ Pauli matrices in a curved space-time by $\sigma^\mu=e^\mu_{\,\,a}\mathbb{S}^a$ and $\bar{\sigma}^\mu=e^\mu_{\,\,a}\bar{\mathbb{S}}^a$. With this definition, the matrices $\gamma^j$ read
\begin{eqnarray}
\label{Weyl gamma matrix j}
\gamma^j&=&\left(
    \begin{matrix}
    0 & -\bar{\sigma}^j\\ 
    {\sigma}^j& 0
    \end{matrix}
    \right).
\end{eqnarray}
Furthermore, observe that the $\sigma^j$ matrices follow the same commutation relations as the flat space-time Pauli matrices. This means that 
$[\sigma^i,\bar\sigma^j]=-\hat{e}^i_k\hat{e}^j_l[\Tilde{\sigma}^k,\Tilde{\sigma}^l]$.
For the Weyl representation we have to obtain two equations for each Dirac fermion. Thus, we need to redefine the covariant derivative $\nabla_\mu$ and the spinor affine connection $\Gamma_\mu$\cite{Poplawski:2007iz}
\cite{Gu:2006qe}, which can be written as $\nabla_\mu=\partial_\mu+\Gamma_\mu$ and $\Gamma_\mu=\dfrac{1}{4}\bar{\sigma}_\nu\sigma^\nu_{;\mu}$, where $\sigma^\mu_{;\nu}=\partial_\nu\sigma^\mu+\Gamma^\mu_{\alpha\nu}\sigma^\alpha$. Nevertheless, in this representation we need to introduce two other notations due to the presence of $\bar{\sigma}^\mu$. Let $\bar{\nabla}_\mu$ and $\Tilde{\Gamma}_\mu$ be the bar covariant derivative and the bar spinor affine connection, respectively, defined by $\bar{\nabla}_\mu=\partial_\mu+\Tilde{\Gamma}_\mu$, where $\Tilde{\Gamma}_\mu=\dfrac{1}{4}\sigma_\nu\bar{\sigma}^\nu_{;\mu}$ (we stress that we use the greek indices for denoting the objects in curved space-time as the gamma and Pauli matrices).

We can now apply the Weyl representation to rewrite the Dirac equation (\ref{Dirac eq}) for a spinor with four components as
\begin{equation}
    \label{weyl eq.1}
    \left( \begin{array}{cc} i\sigma^\mu\left(\bar{\nabla}_\mu+iqA_\mu\right)\psi_R-m\psi_L \\ i\bar{\sigma}^\mu\left(\nabla_\mu+iqA_\mu\right)\psi_L-m\psi_R \end{array} \right)=\left( \begin{array}{cc} 0 \\ 0 \end{array} \right).
\end{equation}
These are the Weyl equations for a spinor in a curved space-time coupled to an electromagnetic field. If we apply the Weyl representation to the transpose conjugated Dirac equation (\ref{transpose conjugated Dirac eq}), it is straightforward to obtain the Weyl equation for the adjoint spinor (\ref{Weyl general adjoint spinor}). However, we shall not write the adjoint spinor equation explicitly because the results are analogous to the spinor equation as we have seen in the previous sections.\newline

 If we set $B=b\Tilde{\gamma}^0$, the current density now reads
\begin{equation}
    \label{Weyl current}
    J^\mu=b\left(\psi_R^\dagger{\sigma}^\mu\psi_R-\psi_L^\dagger{\bar{\sigma}}^\mu\psi_L\right). 
\end{equation}
Explicitly, we have for the spatial part
\begin{eqnarray}
   J^j&=&b\hat{e}^j_i\left(\psi_R^\dagger{\Tilde{\sigma}}^i\psi_R-\psi_L^\dagger\Tilde{\sigma}^i\psi_L+\frac{N^i}{Nb^2}J^0\right).
\end{eqnarray}
On the other hand, the last line of eq. (\ref{eq:diracKG3}) can be obtained from the identities
  \begin{eqnarray}
         \gamma^\mu\gamma^\nu F_{\mu\nu}\psi= \left\{ \begin{array}{cc}  (2NN^kF_{0k}+i\hat{F}_{ij}{\epsilon^{ij}}_k\Tilde{\sigma}^k )\psi_R \\ 
    -(2NN^kF_{0k}-i\hat{F}_{ij}{\epsilon^{ij}}_k\Tilde{\sigma}^k )\psi_L 
    \end{array} \right.,
  \end{eqnarray}
and using definition (\ref{Weyl gamma matrix j}), we find that
\begin{eqnarray}\label{eq:fuentes2}
         \gamma^\mu(\nabla_\mu\gamma^\nu)(D_\nu\psi) &=& \left\{ \begin{array}{cc}  
         -\bar{\mathbb{S}}^a\mathbb{S}^b(\hat\nabla_a\hat{e}^\nu_b)(D_\nu\psi_R) \\ 
    -\mathbb{S}^a\bar{\mathbb{S}}^b(\hat\nabla_a\hat{e}^\nu_b)(D_\nu\psi_L)
    \end{array} \right.\nonumber\\
         &=& \left\{ \begin{array}{cc}  
         (N(\nabla_0 N)-\bar\sigma^j(\nabla_jN))(D_0\psi_R)+(N(\nabla_0\sigma^i)-\bar\sigma^j(\nabla_j\sigma^i))(D_i\psi_R) \\ 
    (N(\nabla_0 N)+\sigma^j(\nabla_jN))(D_0\psi_L)-(N(\nabla_0\bar\sigma^i)-\sigma^j(\nabla_j\bar\sigma^i))(D_i\psi_L)
    \end{array} \right.\nonumber\\
    &=& \left\{ \begin{array}{cc}  
         (\hat\nabla_0 N-\bar{\mathbb{S}}^k(\hat\nabla_kN))(D_0\psi_R)+(\mathbb{S}^k\hat\nabla_0\hat{e}^i_k-\bar{\mathbb{S}}^k\mathbb{S}^l\hat\nabla_k\hat{e}^i_l))(D_i\psi_R) \\ 
    (\hat\nabla_0 N+\mathbb{S}^k(\hat\nabla_kN))(D_0\psi_L)-(\bar{\mathbb{S}}^k\hat\nabla_0\hat{e}^i_k-\mathbb{S}^k\bar{\mathbb{S}}^l(\hat\nabla_k\hat{e}^i_l))(D_i\psi_L),
    \end{array} \right.\nonumber\\
  \end{eqnarray}
where ${\epsilon^{ij}}_k$ is the usual Levi-Civita tensor, $\hat{F}_{ij}=\hat{e}^l_i\hat{e}^m_jF_{lm}$ is the directional Maxwell tensor $\hat{F}_{ij}=(\hat{e}^l_i\hat{\nabla}_j-\hat{e}^l_j\hat{\nabla}_i){A}_l$, and $\hat{\nabla}_a=\hat{e}^\alpha_a\nabla_\alpha$ is the directional covariant derivative which defines the Cartan connection $\hat\nabla_c\hat{e}^\nu_b=\Gamma^a_{bc}\hat{e}_a^\nu$. The Cartan connection $\Gamma^a_{bc}=\hat{e}^a_\nu\hat\nabla_c\hat{e}^\nu_b$ determines the Cartan first fundamental form $d\hat{e}^a+\Gamma^a_b\wedge \hat{e}^b$ for the connections $\Gamma^a_b=\Gamma^a_{bd}\hat{e}^d$ with the property that $\Gamma_{ab}+\Gamma_{ba}=0$, where $\Gamma_{ab}=\eta_{ad}\Gamma^d_b$.\newline

In this section, we have introduced the field equations for Weyl fermions using the relation with the Dirac fermion equations. Moreover, we assume a certain representation for the symmetry Lie gruop to describe the Weyl spinors. In the next section, we will use the field equations found here to get a hydrodynamic representation as in the Dirac spinor case.

 \section{Weyl Hydrodynamic Representation}
\label{section:Weyl Hydrodynamic representation}
We now have all  the ingredients to propose a hydrodynamic representation for the Weyl fermions, following the same procedure as the one developed for the Schrödinger and KG equations in Refs. \cite{Matos:2016ryp,Chavanis:2016shp}.

We start to propose our Madelung transformation in the Weyl spinor, using the exponential map, that is 
\begin{equation}
    \label{weyl hydrodynamic spinor}
    \psi= \left( \begin{array}{cc} \psi_R \\ \psi_L \end{array} \right)=\left( \begin{array}{cc}R_R \\ R_L \end{array} \right)e^{i\theta}.
\end{equation}
Since $\psi_R$ and $\psi_L$ are two spinors, we observe that $R_R$ and $R_L$ are two two-dimensional vectors. The Weyl representation of the adjoint spinor $\bar{\psi}$ when $B=b\Tilde{\gamma}^0$ is 
\begin{align}
    \label{weyl hydro adjoint spinor}
    \bar{\psi}= b\left(  \psi_R^\dagger , \psi_L^\dagger  \right)\Tilde{\gamma}^0&=\left(R_R^\dagger  , R_L^\dagger \right)e^{-i\theta}.
\end{align}
As in section \ref{section: Dirac Hydrodynamic representation}, we use $R_L$ and $R_R$ as complex two-spinors and $\theta$ as a complex function. Therefore, using the Madelung transformation (\ref{weyl hydrodynamic spinor}) in the Weyl equations (\ref{weyl eq.1}) and applying the Lie algebra and the Lie group, we can get the following expression
\begin{equation}
    \label{weyl hydro eq}
    \left( \begin{array}{cc} -\sigma^\mu\left(\bar{\nabla}_\mu  \theta\right)R_R+i\sigma^\mu \left(\bar{\nabla}_\mu R_R\right)-q\sigma^\mu A_\mu R_R \\ -\bar{\sigma}^\mu\left(\nabla_\mu \theta\right)R_L+i\bar{\sigma}^\mu \left(\nabla_\mu R_L\right)-q\bar{\sigma}^\mu A_\mu R_L \end{array} \right)=\left( \begin{array}{cc}m  R_L\\ m  R_R \end{array} \right).
\end{equation}
These are the Weyl equations in curved space-time with the Madelung transformation. We can also apply the Madelung transformation (\ref{weyl hydrodynamic spinor}) and (\ref{weyl hydro adjoint spinor}) to the current density (\ref{Weyl current}), thereby obtaining
\begin{equation}
    \label{Weyl hydro current }
    J^\mu=b\left(R_R^\dagger\bar{\sigma}^\mu  R_R-R_L^\dagger\sigma^\mu R_L\right).
\end{equation}  
 Its components are 
  \begin{eqnarray}\label{Weyl number particle}
       J^0&=&Nb(R_R^\dagger R_R+R_L^\dagger R_L)= Nb n,\\
       J^j&=&b\left(\hat{e}^j_3(n_{\dot 1} - n_{\dot 2} - n_{\dot 3} + n_{\dot 4} )+2\hat{e}^j_1(\sqrt{n_{\dot 1}n_{\dot 2}}-\sqrt{n_{\dot 3}n_{\dot 4}})+\hat{e}^j_i N^i  n\right).\nonumber\\
 \end{eqnarray}
We note that the zero component,
where $ n =\sum_{\dot{\nu}=\dot 1}^{\dot 4} n_{\dot{\nu}}$ is the density number of fermions in the system, gives the number of both right- and left-handed particles.
We can write the following expressions $|\psi_R|^2=\psi^\dagger_R\psi_R=R^\dagger_R R_R= n_R$ and $|\psi_L|^2=\psi^\dagger_L\psi_L=R^\dagger_L R_L= n_L$ for the right- and left-handed spinors, as in the Dirac case. Thus, $ n_R$, $ n_L$ are the right- and left- handed particle number and $n =n_R +n_L$ is the total density number.\newline

Furthermore, eq. (\ref{eq:diracKG3}) using the Weyl representation, which has been discussed in this section, it becomes 
  \begin{eqnarray}\label{eq:diracKGR1}
     i\left[ 2(m v^\mu-\omega\delta^\mu_0)\nabla_\mu R_R-qA_\mu+q\nabla_\mu(A^\mu R_R)+\nabla_\mu(mv^\mu-\omega\delta_0^\mu-qA^\mu)R_R\right]&+&\nonumber\\
     \left(m^2v_{\mu}{v^{\mu}}+2m\omega{v^0}+\frac{\omega^2}{N^2}+m^2\right)R_R-\Box R_R&+&\nonumber\\
     (2NN^kF_{0k}+i{\epsilon^{lj}}_k\hat{F}_{lj}\Tilde{\sigma}^k )R_R&+&\nonumber\\
    (N(\nabla_0 N)-\bar\sigma^j(\nabla_jN))((mv_{0}-\omega)R_R+D_0 R_R)&+&\nonumber\\
    (N(\nabla_0\sigma^k)-\bar\sigma^j(\nabla_j\sigma^k))(imv_{k}R_R+D_kR_R)&=&0.\nonumber\\
\end{eqnarray}   
A similar equation is obtained for the left-handed spinor $R_L$ with the substitution $R\longrightarrow L$  and $\mathbb{S}\longleftrightarrow \bar{\mathbb{S}}$ in eq. (\ref{eq:diracKGR1}). Simplifying the first line in this equation for $\dot\nu=1,2$ corresponding to right-handed components, we get
\begin{eqnarray}\label{eq:diracKGR}
     i\frac{m}{\sqrt{n_{\dot{\nu}}}}\left[-\frac{\omega}{m}\nabla_0n_{\dot{\nu}}+\nabla_\mu(n_{\dot{\nu}}v^\mu)+\frac{\omega}{m}\Box t \right]&+&\nonumber\\
     \sqrt{n_{\dot{\nu}}}\left[m^2v_{\mu}{v^{\mu}}+2m\omega v^0+\frac{\omega^2}{N^2}+m^2-\frac{\Box \sqrt{n_{\dot{\nu}}}}{\sqrt{n_{\dot{\nu}}}}\right]&+&\nonumber\\
     (2NN^kF_{0k}+i{\epsilon^{lj}}_k\hat{F}_{lj}\Tilde{\sigma}^k )R_{R}&+&\nonumber\\
    -(\hat\nabla_a\hat{e}^\alpha_b)\bar{\mathbb{S}}^a\mathbb{S}^b((mv_{\alpha}-\omega\delta^0_\alpha) R_R+D_\alpha R_R)&=&0.\nonumber\\
\end{eqnarray}
The equation for the left-handed components $\dot\nu=3,4$ is obtained by changing $R_R\longrightarrow R_L$ and $\mathbb{S}\longleftrightarrow \bar{\mathbb{S}}$. Note that, although in eq.(\ref{eq:diracKGR}) the first line is multiplied by $i$, we cannot consider the separation between the real and imaginary part, since from the Madelung transformation (\ref{Madelung transformation}) we assume $R$ and $\theta$ as complex parameters. Additionally, the first line of eq. (\ref{eq:diracKGR}) represents the hydrodynamic part of the fermionic fluid. The second line in eq. (\ref{eq:diracKGR}) is written the Bernoulli equation. In this respect, we note that eq. (\ref{hydro dirac eq_b}) is the first integral of this equation. Then, the last lines of eq. (\ref{eq:diracKGR} are the source of the fermionic fluid, something that is not present in the case of bosons. This is because the Dirac equation was introduced \cite{diracequation} in order to eliminate the negative probability problem of the KG equation. As a result, the Dirac equation involves only first derivatives while the KG equation is a second order equation. We will identify the terms in eq.(\ref{eq:diracKGR}) as terms of the first law of thermodynamics  in the next section. \newline

Writing explicitly each component of eq.(\ref{eq:diracKGR}), we can obtain for $\dot{\nu}=\dot{1}$: 
\begin{eqnarray}\label{eq:diracKGRdot1}
     i\frac{m}{\sqrt{n_{\dot{1}}}}\left[-\frac{\omega}{m}\nabla_0n_{\dot{1}}+\nabla_\mu(n_{\dot{1}}v^\mu)+\frac{\omega}{m}\Box t \right]&+&\nonumber\\
     \sqrt{n_{\dot{1}}}\left[m^2v_{\mu}{v^{\mu}}+2m\omega{v^0}+\frac{\omega^2}{N^2}+m^2-\frac{\Box \sqrt{n_{\dot{1}}}}{\sqrt{n_{\dot{1}}}}\right]&=&\nonumber\\
    i\left[ F_{12}\sqrt{n_{\dot{1}}}+F_{23}\sqrt{n_{\dot{2}}}-2\Gamma^a_{21}((m\hat{v}_{a}-\omega\hat{\delta}^0_a) \sqrt{n_{\dot{1}}}+\hat{D}_a \sqrt{n_{\dot{1}}})\right]&-&\nonumber\\
     2i(\Gamma^a_{21}N^1-\Gamma^a_{32}N^3+\Gamma^a_{20}+\Gamma^a_{32})((m\hat{v}_{a}-\omega\hat{\delta}^0_a) \sqrt{n_{\dot{2}}}+\hat{D}_a \sqrt{n_{\dot{2}}})&+&\nonumber\\ 
     2N(F_{01}N^1+F_{02}N^2+F_{03}N^3)\sqrt{n_{\dot{1}}}-F_{13}\sqrt{n_{\dot{2}}}&+&\nonumber\\
     \left[\Gamma^a_{11}(1-({N^1})^2)+\Gamma^a_{22}(1-({N^2})^2)+\Gamma^a_{33}(1-({N^3})^2)\right.&+&\nonumber\\
    \left. 2\Gamma^a_{31}N^1+2\Gamma^a_{32}N^2-\Gamma^a_{00}+2\Gamma^a_{30}\right]((m\hat{v}_{a}-\omega\hat{\delta}^0_a) \sqrt{n_{\dot{1}}}+\hat{D}_a \sqrt{n_{\dot{1}}})&+&\nonumber\\
     (-2\Gamma^a_{21}N^2-2\Gamma^a_{31}N^3+2\Gamma^a_{10}+2\Gamma^a_{31})((m\hat{v}_{a}-\omega\hat{\delta}^0_a) \sqrt{n_{\dot{2}}}+\hat{D}_a \sqrt{n_{\dot{2}}}),&&
\end{eqnarray}
for $\dot{\nu}=\dot{2}$:
\begin{eqnarray}\label{eq:diracKGRdot2}
     i\frac{m}{\sqrt{n_{\dot{2}}}}\left[-\frac{\omega}{m}\nabla_0n_{\dot{2}}+\nabla_\mu(n_{\dot{2}}v^\mu)+\frac{\omega}{m}\Box t \right]&+&\nonumber\\
      \sqrt{n_{\dot{2}}}\left[m^2v_{\mu}{v^{\mu}}+2m\omega{v^0}+\frac{\omega^2}{N^2}+m^2-\frac{\Box \sqrt{n_{\dot{2}}}}{\sqrt{n_{\dot{2}}}}\right]&=&\nonumber\\
     i\left[-F_{12}\sqrt{n_{\dot{2}}}+F_{23}\sqrt{n_{\dot{1}}}+2\Gamma^a_{21}((m\hat{v}_{a}-\omega\hat{\delta}^0_a) \sqrt{n_{\dot{2}}}+\hat{D}_a \sqrt{n_{\dot{2}}})\right]&+&\nonumber\\
     2i(\Gamma^a_{21}N^1-\Gamma^a_{32}N^3+\Gamma^a_{20}-\Gamma^a_{32})(m\hat{v}_{a}-\omega\hat{\delta}^0_a) \sqrt{n_{\dot{1}}}+\hat{D}_a \sqrt{n_{\dot{1}}})&+&\nonumber\\
     2N(F_{01}N^1+F_{02}N^2+F_{03}N^3)\sqrt{n_{\dot{2}}}+F_{13}\sqrt{n_{\dot{1}}}&+&\nonumber\\
     \left[\Gamma^a_{11}(1-({N^1})^2)+\Gamma^a_{22}(1-({N^2})^2)+\Gamma^a_{33}(1-({N^3})^2\right.)&+&\nonumber\\
     \left.-2\Gamma^a_{31}N^1-2\Gamma^a_{32}N^2-\Gamma^a_{00}-2\Gamma^a_{30}\right]((m\hat{v}_{a}-\omega\hat{\delta}^0_a) \sqrt{n_{\dot{2}}}+\hat{D}_a \sqrt{n_{\dot{2}}})&+&\nonumber\\
     (-2\Gamma^a_{21}N^2-2\Gamma^a_{31}N^3+2\Gamma^a_{10}-2\Gamma^a_{31})((m\hat{v}_{a}-\omega\hat{\delta}^0_a) \sqrt{n_{\dot{1}}}+\hat{D}_a \sqrt{n_{\dot{1}}}),&&
\end{eqnarray}
for $\dot{\nu}=\dot{3}$:
\begin{eqnarray}\label{eq:diracKGRdot3}
     i\frac{m}{\sqrt{n_{\dot{3}}}}\left[-\frac{\omega}{m}\nabla_0n_{\dot{3}}+\nabla_\mu(n_{\dot{3}}v^\mu)+\frac{\omega}{m}\Box t \right]&+&\nonumber\\
     \sqrt{n_{\dot{3}}}\left[m^2v_{\mu}{v^{\mu}}+2m\omega{v^0}+\frac{\omega^2}{N^2}+m^2-\frac{\Box \sqrt{n_{\dot{3}}}}{\sqrt{n_{\dot{3}}}}\right]&=&\nonumber\\
     i\left[F_{12}\sqrt{n_{\dot{3}}}+F_{23}\sqrt{n_{\dot{4}}}-2\Gamma^a_{21}((m\hat{v}_{a}-\omega\hat{\delta}^0_a) \sqrt{n_{\dot{3}}}+\hat{D}_a \sqrt{n_{\dot{3}}})\right]&+&\nonumber\\
     2i(\Gamma^a_{21}N^1-\Gamma^a_{32}N^3+\Gamma^a_{20}-\Gamma^a_{32})((m\hat{v}_{a}-\omega\hat{\delta}^0_a) \sqrt{n_{\dot{4}}}+\hat{D}_a \sqrt{n_{\dot{4}}})&+&\nonumber\\
     2N(F_{01}N^1+F_{02}N^2+F_{03}N^3)\sqrt{n_{\dot{3}}}-F_{13}\sqrt{n_{\dot{4}}}&+&\nonumber\\
     \left[\Gamma^a_{11}(1-({N^1})^2)+\Gamma^a_{22}(1-({N^2})^2)+\Gamma^a_{33}(1-({N^3})^2)\right.&+&\nonumber\\
     \left.-2\Gamma^a_{31}N^1-2\Gamma^a_{32}N^2-\Gamma^a_{00}-2\Gamma^a_{30}\right]((m\hat{v}_{a}-\omega\hat{\delta}^0_a) \sqrt{n_{\dot{3}}}+\hat{D}_a \sqrt{n_{\dot{3}}})&+&\nonumber\\
     (2\Gamma^a_{21}N^2+2\Gamma^a_{31}N^3-2\Gamma^a_{10}+2\Gamma^a_{31})((m\hat{v}_{a}-\omega\hat{\delta}^0_a) \sqrt{n_{\dot{4}}}+\hat{D}_a \sqrt{n_{\dot{4}}}),&&
\end{eqnarray}
and for $\dot{\nu}=\dot{4}$:
\begin{eqnarray}\label{eq:diracKGRdot4}
     i\frac{m}{\sqrt{n_{\dot{4}}}}\left[-\frac{\omega}{m}\nabla_0n_{\dot{4}}+\nabla_\mu(n_{\dot{4}}v^\mu)+\frac{\omega}{m}\Box t \right]&+&\nonumber\\
     \sqrt{n_{\dot{4}}}\left[m^2v_{\mu}{v^{\mu}}+2m\omega{v^0}+\frac{\omega^2}{N^2}+m^2-\frac{\Box \sqrt{n_{\dot{4}}}}{\sqrt{n_{\dot{4}}}}\right]&=&\nonumber\\
    i\left[-F_{12}\sqrt{n_{\dot{4}}}+F_{23}\sqrt{n_{\dot{3}}}+2\Gamma^a_{21}((m\hat{v}_{a}-\omega\hat{\delta}^0_a) \sqrt{n_{\dot{4}}}+\hat{D}_a \sqrt{n_{\dot{4}}})\right]&-&\nonumber\\
     2i(\Gamma^a_{21}N^1-\Gamma^a_{32}N^3+\Gamma^a_{20}+\Gamma^a_{32})((m\hat{v}_{a}-\omega\hat{\delta}^0_a) \sqrt{n_{\dot{3}}}+\hat{D}_a \sqrt{n_{\dot{3}}})&+&\nonumber\\
     2N(F_{01}N^1+F_{02}N^2+F_{03}N^3)\sqrt{n_{\dot{4}}}+F_{13}\sqrt{n_{\dot{3}}}&+&\nonumber\\
     \left[\Gamma^a_{11}(1-({N^1})^2)+\Gamma^a_{22}(1-({N^2})^2)+\Gamma^a_{33}(1-({N^3})^2)\right.&+&\nonumber\\
     \left.2\Gamma^a_{31}N^1+2\Gamma^a_{32}N^2-\Gamma^a_{00}+2\Gamma^a_{30}\right]((m\hat{v}_{a}-\omega\hat{\delta}^0_a) \sqrt{n_{\dot{4}}}+\hat{D}_a \sqrt{n_{\dot{4}}})&+&\nonumber\\
     (2\Gamma^a_{21}N^2+2\Gamma^a_{31}N^3-2\Gamma^a_{10}-2\Gamma^a_{31})((m\hat{v}_{a}-\omega\hat{\delta}^0_a) \sqrt{n_{\dot{3}}}+\hat{D}_a \sqrt{n_{\dot{3}}}),&&
\end{eqnarray}
where we have used that $\Gamma_{ab}+\Gamma_{ba}=0$ and defined the directional quantities $v_{a}=v_{\alpha}\hat{e}_a^\alpha$, $\hat{\delta}^0_a=\delta^0_\alpha \hat{e}_a^\alpha=N\delta^0_a$ and $\hat{D}_a=\hat{e}_a^\alpha D_\alpha $.
\newline

Observe that the structure of equations (\ref{eq:diracKGRdot1})-(\ref{eq:diracKGRdot4}) is
\begin{eqnarray}\label{eq:diracKGRdot4bb}
     i\frac{m}{\sqrt{n_{\dot{\nu}}}}\left[-\frac{\omega}{m}\nabla_0n_{\dot{\nu}}+\nabla_\mu(n_{\dot{\nu}}v^\mu)+\frac{\omega}{m}\Box t \right]&+&\nonumber\\
      \sqrt{n_{\dot{\nu}}}\left[m^2v_{\mu}{v^{\mu}}+2m\omega{v^0}+\frac{\omega^2}{N^2}+m^2-\frac{\Box \sqrt{n_{\dot{\nu}}}}{\sqrt{n_{\dot{\nu}}}}\right]&=&\nonumber\\
     i\left[e_{1\dot{\nu}}F_{12}\sqrt{n_{\dot{\nu}}}+F_{23}\sqrt{n_{\ddot{\nu}}}-2e_{1\dot{\nu}}\Gamma^a_{21}((m\hat{v}_{a}-\omega\hat{\delta}^0_a) \sqrt{n_{\dot{\nu}}}+\hat{D}_a \sqrt{n_{\dot{\nu}}})\right]&-&\nonumber\\
     2i(\Gamma^a_{21}N^1-\Gamma^a_{32}N^3+\Gamma^a_{20}+e_{2\dot{\nu}}\Gamma^a_{32})((m\hat{v}_{a}-\omega\hat{\delta}^0_a) \sqrt{n_{\ddot{\nu}}}+\hat{D}_a \sqrt{n_{\ddot{\nu}}})&+&\nonumber\\
    2N(F_{01}N^1+F_{02}N^2+F_{03}N^3)\sqrt{n_{\dot{\nu}}}-e_{1\dot{\nu}}F_{13}\sqrt{n_{\ddot{\nu}}}&+&\nonumber\\
     \left[\Gamma^a_{11}(1-({N^1})^2)+\Gamma^a_{22}(1-({N^2})^2)+\Gamma^a_{33}(1-({N^3})^2)\right.&+&\nonumber\\
     \left.2e_{2\dot{\nu}}(\Gamma^a_{31}N^1+\Gamma^a_{32}N^2+\Gamma^a_{30})-\Gamma^a_{00}\right]((m\hat{v}_{a}-\omega\hat{\delta}^0_a) \sqrt{n_{\dot{\nu}}}+\hat{D}_a \sqrt{n_{\dot{\nu}}})&+&\nonumber\\
     (-2e_{3\dot{\nu}}(\Gamma^a_{21}N^2+\Gamma^a_{31}N^3-\Gamma^a_{10})+2e_{1\dot{\nu}}\Gamma^a_{31})((m\hat{v}_{a}-\omega\hat{\delta}^0_a) \sqrt{n_{\ddot{\nu}}}+\hat{D}_a \sqrt{n_{\ddot{\nu}}}),&&
\end{eqnarray}
where the coefficients $e_{i\dot{\nu}}$ are $\pm 1$ with $e_{1\dot{\nu}}=(+,-,+,-)$, $e_{2\dot{\nu}}=(-,+,-,+)$ and $e_{3\dot{\nu}}=(+,+,-,-)$, and the sub-index $\ddot{\nu}$ are the conjugate of the sub-index $\dot{\nu}$, such that $\ddot{1}=\dot{2}$, $\ddot{2}=\dot{1}$, $\ddot{3}=\dot{4}$ and $\ddot{4}=\dot{3}$. In comparison with the boson case, we cannot separate them in real and imaginary part. Due to, the generalized transformation, that we assume, has complex parameters. Therefore, we shall work with the full equations, which are more complicated than the standard equations for fermions in curved space-time, that themselves are complicated. An advantage for the hydrodynamic representation, that we found, is to give directly an interpretation of quantum theory through the De Broglie-Bohm interpretation.\newline

\section{Energy Balance}
\label{section: energy balance}

From equation (\ref{eq:diracKGR}), we can identify the different energy contributions to the Fermi gas, and obtain an   energy balance equation for fermions analogous to the one obtained for bosons in  \cite{Matos:2016ryp,Chavanis:2016shp}. In order to simplify the notations, we can re-write the equation (\ref{eq:diracKGR}) in terms of the $\dot{\nu}$ coefficients with the understanding that the subindex $R$ refers to each component $R=\dot{1},\dot{2}$ individually. We get
\begin{eqnarray}
\label{eq:balance_energy}
     i\left[-\omega\nabla_0\ln(n_{\dot{\nu}})+\dfrac{m\nabla_\mu( n_{\dot{\nu}}v^\mu)}{n_{\dot{\nu}}}+\frac{\omega}{n_{\dot{\nu}}}\Box t \right]&+&\nonumber\\
    2m^2\left(K+\frac{1}{m}\omega{v^0}+\frac{1}{2}U^N+U^Q\right)+ E+U^S&=&0.
\end{eqnarray}
%This equation is valid for right handed fermions. The result is the same for left handed fermions changing $R_R\longrightarrow R_L$ in the first line, and $\mathbb{S}\longleftrightarrow \bar{\mathbb{S}}$ in the second line. \newline
The first line in eq. (\ref{eq:balance_energy}) describes the free density evolution of the fermions, while the contribution of the different energy terms appears in the second line. The first one is the kinetic energy $K_{\dot{\nu}}$ defined as 
\begin{eqnarray}
K=\dfrac{1}{2}v_{\mu}v^\mu.
\end{eqnarray}
The lapse potential $U^N$ is given by
\begin{eqnarray}
\label{eq:lapse potential term}
U^N=\frac{\omega^2}{m^2}\frac{1}{N^2}+1.
\end{eqnarray}
It represents the energy contribution due to the chosen lapse function $N$. The quantum potential $U^Q$ is defined as
\begin{equation}
U^Q=-\frac{1}{2m^2}\dfrac{\Box \sqrt{n_{\dot{\nu}}}}{\sqrt{n_{\dot{\nu}}}}.    
\end{equation}
The contribution of the electromagnetic interaction $E$ is given by 
\begin{eqnarray}
\label{eq:electromagn potential term}
   E&=& (2NN^kF_{0k}+i{\epsilon^{lj}}_k\hat{F}_{lj}\Tilde{\sigma}^k), \\
   &=&2N(F_{01}N^1+F_{02}N^2+F_{03}N^3)-e_{1\dot{\nu}}F_{13}\sqrt{\frac{n_{\ddot{\nu}}}{n_{\dot{\nu}}}}+i\left(e_{1\dot{\nu}}F_{12}+F_{23}\sqrt{\frac{n_{\ddot{\nu}}}{n_{\dot{\nu}}}}\right).
   \nonumber
\end{eqnarray}
It depends on the Faraday tensor, shift vector and lapse function that are related to the Pauli matrices. This relationship is due to the interaction between the electromagnetic field and the fermionic spin. Finally, the potential $U^S_{\dot{\nu}}$ describes the interaction between the spin and the geometry of space-time. It is given by 
\begin{eqnarray}
\label{eq:spin-geom potential term}
    U^S&=&-\left((m\hat{v}_{Rd}-\omega_{\dot{\nu}}\hat{\delta}^0_d) +\dfrac{\hat{D}_\alpha\sqrt{n_{\dot{\nu}}}}{\sqrt{n_{\dot{\nu}}}}\right)\Gamma^d_{ba}\bar{\mathbb{S}}^a\mathbb{S}^b,\\
    &=&\left[\Gamma^a_{11}(1-({N^1})^2)+\Gamma^a_{22}(1-({N^2})^2)+\Gamma^a_{33}(1-({N^3})^2)\right.\nonumber\\
     &+&\left.2e_{2\dot{\nu}}(\Gamma^a_{31}N^1+\Gamma^a_{32}N^2+\Gamma^a_{30})-\Gamma^a_{00}\right]\left((m\hat{v}_{a}-\omega\hat{\delta}^0_a)+\frac{\hat{D}_a \sqrt{n_{\dot{\nu}}}}{\sqrt{n_{\dot{\nu}}}}\right)\nonumber\\
     &+&(-2e_{3\dot{\nu}}(\Gamma^a_{21}N^2+\Gamma^a_{31}N^3-\Gamma^a_{10})+2e_{1\dot{\nu}}\Gamma^a_{31})\left((m\hat{v}_{a}-\omega\hat{\delta}^0_a) \sqrt{\frac{n_{\ddot{\nu}}}{n_{\dot{\nu}}}}+\frac{\hat{D}_a \sqrt{n_{\ddot{\nu}}}}{ \sqrt{n_{\dot{\nu}}}}\right)\nonumber\\
     &+&i\left[-2e_{1\dot{\nu}}\Gamma^a_{21}\left((m\hat{v}_{a}-\omega\hat{\delta}^0_a) +\frac{\hat{D}_a \sqrt{n_{\dot{\nu}}}}{ \sqrt{n_{\dot{\nu}}}}\right)\right.
     \nonumber\\
     &-&\left.2(\Gamma^a_{21}N^1-\Gamma^a_{32}N^3+\Gamma^a_{20}+e_{2\dot{\nu}}\Gamma^a_{32})\left((m\hat{v}_{a}-\omega\hat{\delta}^0_a) \sqrt{\frac{n_{\ddot{\nu}}}{n_{\dot{\nu}}}}+\frac{\hat{D}_a \sqrt{n_{\ddot{\nu}}}}{ \sqrt{n_{\dot{\nu}}}}\right)\right],
\end{eqnarray}
%for $\dot{\nu}=\dot{1},\dot{2}$, and by making the substitution $\mathbb{S}\longleftrightarrow \bar{\mathbb{S}}$ for $\dot{\nu}=\dot{3},\dot{4}$.
%In the foregoing equations, the notation $|_{\dot{\nu}}$ means that we have to evaluate the quantity at the corresponding $\dot{\nu}$.
Note that $U^S$ disappears if we assume a flat space-time or if we consider particles without spin. Furthermore, $U^S$ is constructed with the generalized gamma matrices (\ref{Weyl gamma matrix j}), which are related to the spin (the Pauli matrices) and to the space-time geometry (tetrads).\newline

Finally, we can also write equation (\ref{eq:diracKG2}) as a Gross-Pitaevskii-like equation. If we perform the transformation $\psi=\Psi e^{i\omega_0 t}$, where $\Psi$ is a four spinor that depends on all the variables $x^\mu$, equation (\ref{eq:diracKG2}) becomes
\begin{eqnarray}\label{eq:Schrodinger}
     i\nabla^0\Psi-\frac{1}{2\omega_0}\Box_E\Psi+\frac{m^2}{2\omega_0}\Psi+\left(-\frac{\omega_0}{N^2}-2qA^0+i\Box t\right)\Psi&+&\nonumber\\
      \frac{1}{2\omega_0}\left(
    \begin{matrix}
    2NN^kF_{0k}+i\hat{F}_{ij}{\epsilon^{ij}}_k\Tilde{\sigma}^k  & 0\\ 
    0 & -2NN^kF_{0k}+i\hat{F}_{ij}{\epsilon^{ij}}_k\Tilde{\sigma}^k
    \end{matrix}
    \right)\Psi&-&\nonumber\\
    \frac{1}{2\omega_0}\left(
   \begin{matrix}
    \bar{\mathbb{S}}^a\mathbb{S}^b & 0\\ 
    0 & \mathbb{S}^a\bar{\mathbb{S}}^b
    \end{matrix}
    \right)\Gamma^d_{ba}(\hat{D}_d\Psi+i\omega_0N\delta^0_d\Psi)&=&0.
\end{eqnarray}
Equation (\ref{eq:Schrodinger}) is the generalization of the Gross-Pitaevskii equation\cite{Gross1961StructureOA} for fermions with electromagnetic field interaction in an arbitrary space-time.\newline

\section{Conclusions}
\label{section: conlusions}

A non-standard representation for fermions was worked using an analogy as in the boson and quantum mechanics case, where it was proposed the Madelung transformation. We extended this transformation for the spinor case, either Dirac or Weyl fermions. Thus, it was possible to get a successful hydrodynamic representation for fermions in an arbitrary framework coupled to an electromagnetic field. Although, the full equations that describe the Fermi gas behaviour are more complicated than in standard description. This is closer to the De Broglie-Bohm interpretation in quantum theory, where the measure problem can be solved by a statistic way. Furthermore, a non-obvious result using this new description was the first law of the thermodynamics or the energy balance equation, where different energy contributions of these kind of particles were found.\newline

The main difference between the hydrodynamic representation of bosons \cite{Matos:2016ryp}\cite{Chavanis:2016shp} and fermions, concerns the form of the Bernoulli equation. For bosons, after doing the Madelung transformation, we can separate the KG equation into real and imaginary parts. By contrast, for fermion particles we have to work with the complete equations of motion because the  real and imaginary parts cannot be easily separated. This is related to the fact that the gamma matrices are a  representation of the SO$(1,3)$ group and the generalized Madelung transformation used, because it only admits complex parameter to fulfill the Lorentz invariance.\newline

The spin is a fundamental outcome of the Dirac equation \cite{diracequation}, which combines both elements of special relativity and quantum mechanics, that was introduced to solve the problem of negative probability present in the KG equation -- first proposed as a relativistic generalization of the Schr\"odinger equation. Here, we observe that the general relativistic Dirac equation involves an additional contribution due to geometry and spin through the generalized gamma and Pauli matrices. These terms arise from endowing a quantum field with a curvature (geometry) given by a metric in General Relativity. Such a  contribution is absent  in a flat space-time and in a system without spin as for a scalar field.\newline

With this work we open the possibility of studying in detail the behavior of fermions in different situations (such as massive stars or dark matter halos harboring a central black hole), where general relativity effects may be important. We solved the problem of energy balance for both bosons and fermions. In this manner, we can compare the result of the hydrodynamic representation for classical and quantum fluids in the various geometries mentioned above.

\section{Acknowledgments}
\label{section:Acknowledgments}

This work was partially supported by CONACyT M\'exico under grants: A1-S-8742, 304001, 376127; Project No. 269652 and Fronteras Project 281;Xiuhcoatl and Abacus clusters at Cinvestav, IPN; I0101/131/07 C-234/07 of the Instituto Avanzado de Cosmolog\'ia (IAC) collaboration (http:// www.iac.edu.mx). O.G. acknowledges financial support from CONACyT doctoral fellowship and appreaciates Angelica C. Aguirre Castañ\'on for her valuable review and support. Works of T.M. are partially supported by Conacyt through the Fondo Sectorial de Investigaci\'on para la Educaci\'on, grant CB-2014-1, No. 240512.

\bibliographystyle{ieeetr}
\bibliography{ref.bib}

\appendix

\section{Solutions to the Dirac equation in flat space-time}
\label{appendixA}
Equation (\ref{Dirac eq}) in flat space-time, using the Pauli matrices (\ref{eq:Pauli}), reads
\begin{eqnarray}\label{eq:DiracExpl}
\left[ \begin {array}{c} {\frac {\partial }{\partial t}}\psi_y 
-{\frac {\partial }{\partial x}}\psi_z 
+i{\frac {\partial }{\partial y}}\psi_z 
-{\frac {\partial }{\partial z}}\psi_y -m\psi_t 
\\ \noalign{\medskip}{\frac {\partial }{\partial t}}\psi_z 
-{\frac {\partial }{\partial x}}\psi_y 
-i{\frac {\partial }{\partial y}}\psi_y 
+{\frac {\partial }{\partial z}}\psi_z 
-m\psi_x \\ \noalign{\medskip}{\frac {\partial }{\partial t}}\psi_t 
 +{\frac {\partial }{\partial x}}\psi_x -i{\frac {\partial }{\partial y}}\psi_x 
 +{\frac {\partial }{\partial z}}\psi_t 
 -m\psi_y \\ 
 \noalign{\medskip}{\frac {\partial }{\partial t}}\psi_x 
 +{\frac {\partial }{\partial x}}\psi_t 
 +i{\frac {\partial }{\partial y}}\psi_t 
 -{\frac {\partial }{\partial z}}\psi_x 
 -m\psi_z \end {array} \right] =0,
\end{eqnarray}
  where we have defined the spinor as $\psi=(\psi_{\dot\mu})=(\psi_x,\psi_y,\psi_z,\psi_t)^T$. In order to find an exact solution of the previous equation, we use the ansatz $\psi_{\dot\mu}=R_{0\dot\mu}\exp(i(x_0 x+y_0 y+z_0 z + t_0 t))$, where $x_0\cdots t_0$ and $R_{0\dot\mu}$ are constants. Here, we have the simplest solutions of the Dirac equation where the exponential is the same for all components. We obtain four linear equations
  \begin{eqnarray}
  iR_{0z}\zeta_0^*+iR_{0y}\eta_0+mR_{0t}&=&0,\nonumber\\
  iR_{0y}\zeta_0-iR_{0z}\xi_0+mR_{0x}&=&0,\nonumber\\
  R_{0x}\zeta_0^*+R_{0t}\xi_0+imR_{0y}&=&0,\nonumber\\
  R_{0t}\zeta_0-R_{0x}\eta_0+imR_{0z}&=&0,
  \end{eqnarray}
where $\zeta_0=x_0+iy_0$, $\eta_0=z_0-t_0$, and $\xi_0=z_0+t_0$. The solutions of these equations are
\begin{eqnarray}
R_{0t}&=&-\frac{1}{m}(iR_{0y}\eta_0+iR_{0z}\zeta_0),\nonumber\\
R_{0x}&=&\frac{1}{m}(iR_{0z}\xi_0-iR_{0y}\zeta_0^*),
\end{eqnarray}
where $x_0^2+y_0^2+z_0^2-t_0^2=m^2$. 

Now, we use the ansatz $\psi_\mu=R_{0\mu}\exp(i\theta)$, where $\theta$ is an arbitrary function of the coordinates. Substituting this ansatz into (\ref{eq:DiracExpl}), we obtain
\begin{eqnarray}
iR_{0z}Z_0^*+iR_{0y}E_0+mR_{0t}&=&0,\nonumber\\
  iR_{0y}Z_0-iR_{0z}F_0+mR_{0x}&=&0,\nonumber\\
  R_{0x}Z_0^*+R_{0t}F_0+imR_{0y}&=&0,\nonumber\\
  R_{0t}Z_0-R_{0x}E_0+imR_{0z}&=&0,
\end{eqnarray}
where $Z_0=\theta_{,x}+i\theta_{,y}$, $E_0=\theta_{,z}-\theta_{,t}$, and $F_0=\theta_{,z}+\theta_{,t}$. The solution of the previous system of differential equations is
\begin{eqnarray}
\theta&=&F(X)-it\nonumber\\
&+&\frac{m}{2R_{0t}R_{0z}+2R_{0x}R_{0y}}\left(i\zeta_0^*(R_{0x}^2-R_{0z}^2)
-i\zeta_0 (R_{0y}^2-R_{0t}^2)\right),
\end{eqnarray}
where $F(X)$ is an arbitrary function of
\begin{equation}
    X=\frac{R_{0t}(-\zeta R_{0y}-\zeta^*R_{0x}+\xi R_{0y}-\eta R_{0z})}{2R_{0t}R_{0z}+2R_{0x}R_{0y}}.
\end{equation}

\end{document}